\documentclass[prc,showpacs,showkeys]{revtex4}
\usepackage{epsfig,graphicx,float,color}
\usepackage{amssymb}
\usepackage[normalem]{ulem}

\begin{document}

\title{The platinum nuclei: concealed configuration mixing and shape
  coexistence}
\author{J.E. Garc\'{\i}a-Ramos$^1$, V. Hellemans$^{2,3}$, K. Heyde$^4$ } 
\affiliation{
$^1$Departamento de F\'{\i}sica Aplicada, Universidad de Huelva,
21071 Huelva, Spain\\
$^2$Universit\'e Libre de Bruxelles, Physique Nucl\'eaire Th\'eorique
et Physique Math\'ematique, CP229, B-1050 Brussels, Belgium\\
$^{3}$University of Notre Dame, Department of Physics, 225 
Nieuwland Science Hall, Notre Dame IN 46556, USA\\
$^4$Department of Physics and Astronomy, Proeftuinstraat, 86 B-9000 Ghent, Belgium\\
}

\begin{abstract}
The role of configuration mixing in the Pt region is investigated. For
this chain of isotopes, the nature of the ground state changes smoothly, 
being spherical around mass $A\sim 174$ and $A\sim 192$ and 
deformed around the mid-shell $N=104$ region. This has a
dramatic effect on the systematics of the energy spectra as compared to
the systematics in the Pb and Hg nuclei. Interacting Boson Model with configuration
mixing calculations are presented for gyromagnetic factors, $\alpha$-decay
hindrance factors, and isotope shifts. The necessity of incorporating intruder
configurations to obtain an accurate description of the latter properties becomes 
evident. 
\end{abstract}
\pacs{21.10.-k, 21.60.-n, 21.60.Fw.}

\keywords{Pt isotopes, shape coexistence, intruder states.}
\maketitle

\section{Introduction}
\label{sec-intro}
By now, shape coexistence has been observed in many mass regions
throughout the nuclear chart and turns out to be realized in
more nuclei than anticipated a few decades ago \cite{heyde11}.  
Shell-model \cite{caurier05} and mean-field \cite{bender03} approaches
have shown that shape coexistence arises naturally, in the first case through inclusion of
many-particle, many-hole excitations across closed shells and in the latter case through constraints on
the quadrupole moment in Hartree Fock (HF) and Hartree-Fock-Bogoliubov (HFB)
studies \cite{grahn08,duguet03,smirnova03,bender04,rodri10,nomura11}. A particularly well-documented 
example of shape coexistence is the Pb region. From the closed neutron
shell ($N=126$) to the very neutron-deficient nuclei, approaching and
even going beyond the $N=104$ mid-shell, ample experimental evidence
for shape coexisting bands has been accumulated for the Pb ($Z=82$) and
Hg ($Z=80$) nuclei \cite{jul01,hey83,wood92}.  Recent experiments have
extended our knowledge of the excitation energies in intruding bands \cite{rahkila10},
lifetime data \cite{grahn06,grahn08,grahn09a,grahn09,scheck10}, nuclear charge radii
\cite{dewitte07,cocio11}, 2$_1^+$ gyromagnetic factors
\cite{stuchbery96,bian07}, and $\alpha$-decay hindrance factors
\cite{wauters94,wauters94a,duppen00}.  

\begin{figure}[hbt]
  \centering
  \epsfig{file=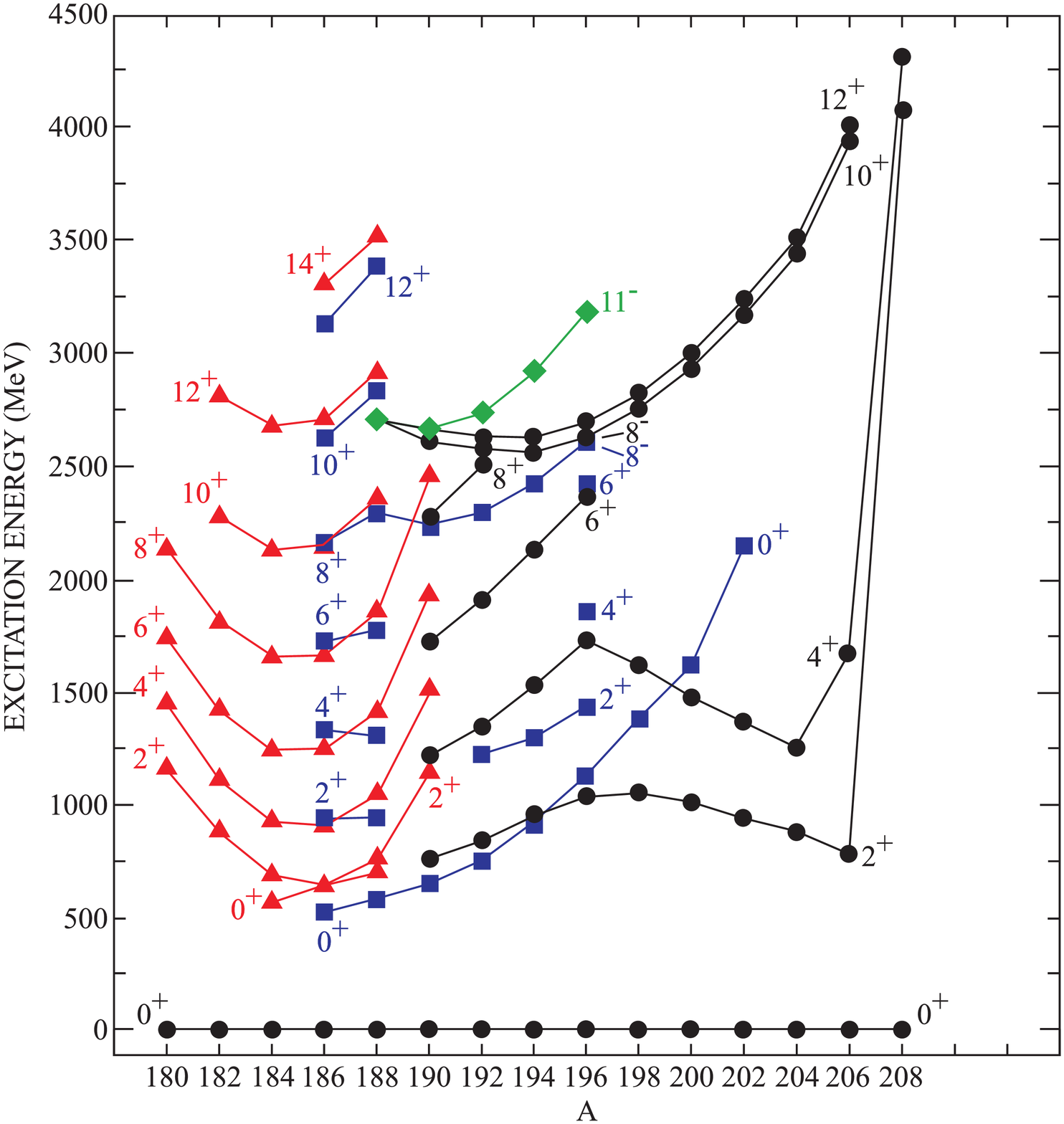,width=10cm} 
  \caption{(Color online) Energy systematic of the Pb nuclei. The red full lines
    connect states associated with a prolate structure, the 
  blue dashed lines states with an oblate structure and the black
  lines connect states with a spherical structure.} 
  \label{fig-Pb-syst}
\end{figure}

\begin{figure}[hbt]
  \centering
  \epsfig{file=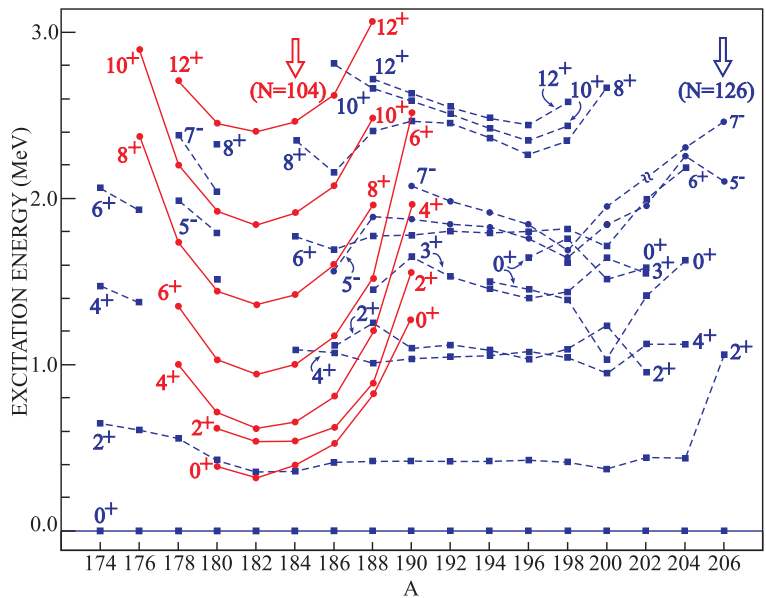,width=10cm} 
  \caption{(Color online) Energy systematic of the Hg nuclei. The red lines connect states associated with a prolate structure and the
  blue lines states with an oblate structure.}
  \label{fig-Hg-syst}
\end{figure}

\begin{figure}[hbt]
  \centering
  \epsfig{file=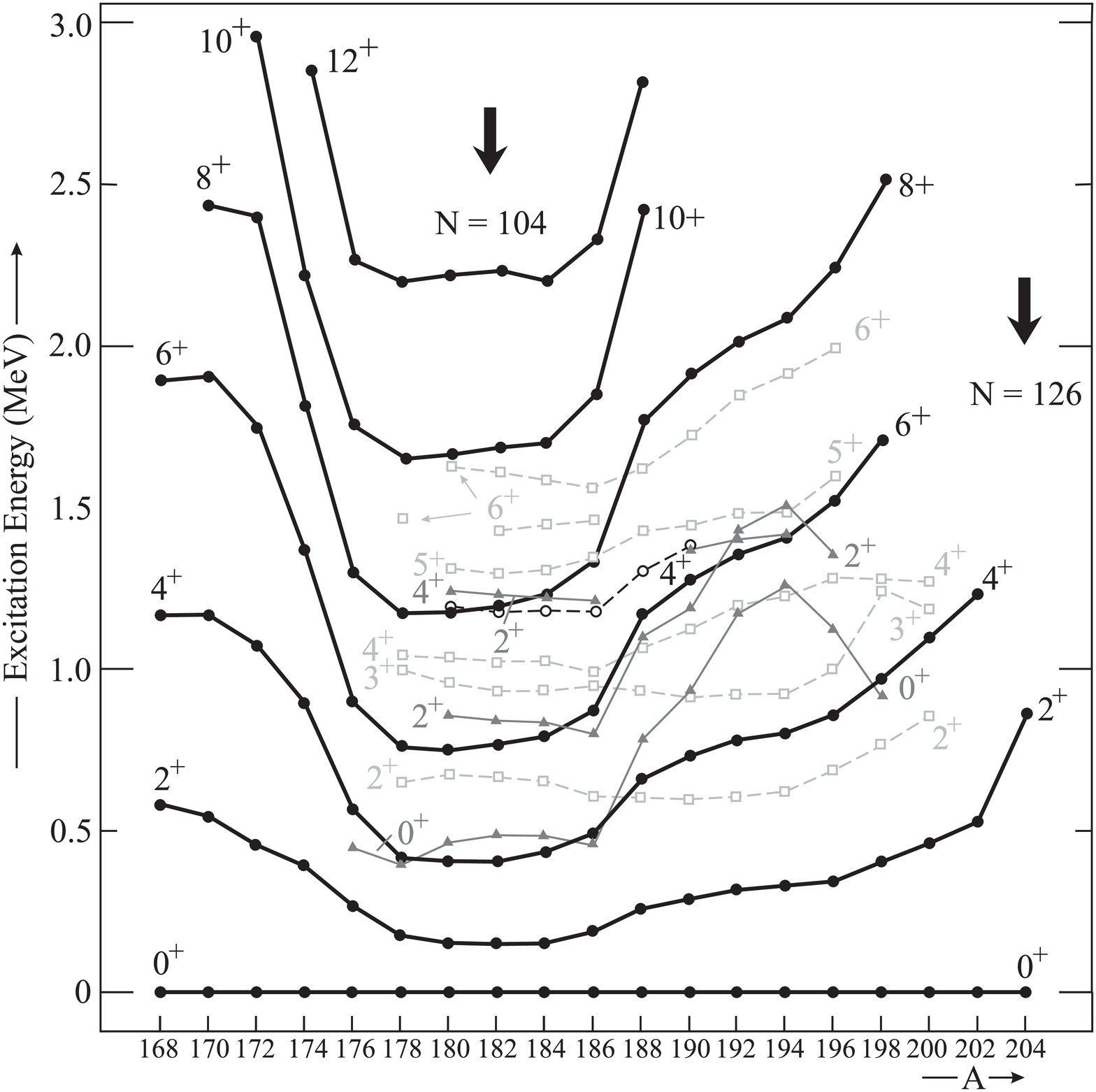,width=10cm} 
  \caption{Energy systematic of the Pt nuclei. The dark lines connect
    the yrast band structure, the full and dashed thin lines connect the
    non-yrast levels.} 
  \label{fig-Pt-syt}
\end{figure}

An important question is how these shape coexisting structures will
evolve when one moves further away from the $Z=82$ and $N=126$ closed
shells. Recently, a lot of new results have become available
for the even-even Po, Hg and Pt nuclei, for which experimental information was 
highly needed. It is informative to compare the systematics of the low-lying states
of the $Z=82$ proton closed shell Pb nuclei (Fig.~\ref{fig-Pb-syst}),
the $Z=80$ Hg nuclei (Fig.~\ref{fig-Hg-syst}), and the $Z=78$ Pt nuclei
(Fig.~\ref{fig-Pt-syt}). 
The data to construct these figures have been taken from the relevant Nuclear
Data Sheets, from \cite{rahkila10} (for the Pb nuclei), from 
\cite{sand09,bree10,huyse10,scheck10,scheck11} (for the Hg nuclei), and 
\cite{cutcham05,williams06,joss06,oktem07,cutcham08,gomez09,ilie10} (for the Pt nuclei).
Whereas the intruder bands are easily singled out  for the Pb and Hg nuclei 
and the excitation energies display the characteristic parabolic pattern with
minimal excitation energy around the $N=104$ neutron mid-shell nucleus, this
structure seems lost for the Pt nuclei. Focussing on the systematics of the
energy spectra in these Pt nuclei as a function of the neutron number,
one observes a rather sudden drop in the excitation energy of the 0$^+_2$, 4$^+_1$, 
2$^+_3$ and 6$^+_1$ states between $N=110$ ($A=188$)
and $N=108$ ($A=186$), followed by a particularly flat behaviour as a function
of $N$ until the energies of those states start to move up again around neutron 
number $N=100$ ($A=178$).

As the Pb nuclei, the Pt nuclei have been studied within the
framework of the Interacting Boson Model (IBM) \cite{iach87}. Taking into
account the presence of proton 2p-2h excitations across the $Z=82$
proton closed shell \cite{duval82}, one achieves an overall good description of both energy
spectra, radii, and g-factors \cite{harder97,king98}. In addition,
IBM calculations that do not explicitly take into account the
proton intruder configurations have also been carried
out \cite{cutcham05,cutcham05a,cutcham08}, resulting in a
satisfactory description of excitation energies and $B(E2)$ transition
rates.  In a previous paper \cite{Garc09}, we studied the Pt nuclei extensively and
carried out a detailed comparison between calculations that include proton
2p-2h excitations (hence, in the model space $[N]
\oplus [N+2]$, where $N$ denotes the total number of bosons, irrespective
of their charge character) with calculations that consider the smaller model
space of the $[N]$ configurations only. It turned out that the results for the
energy spectra and absolute $B(E2)$ values were very similar up
to an excitation energy of $\sim$ 1.5 MeV, even though the
corresponding wave functions have to be very
different in some cases. As such, it was concluded that these
similarities point towards a picture where the configuration mixing
and the larger model space are somehow ``concealed''.

This very same observation has been put forward a long time ago by
Cohen, Lawson and Soper \cite{cohen66,law67,law80} when
addressing the question ``How can the results using a large model
space, be very similar to the ones resulting from a truncated model
space''.
Starting from a model space of two degenerate $1d_{3/2}$ and $1f_{7/2}$ single-particle neutron orbitals 
(containing a neutron number ranging from 4 to 12) and a given two-body interaction, a Yukawa potential, 
they constructed a set of theoretical nuclei which were called the Pseudonium nuclei  $^{40-48}$Ps. 
Interpreting the Ps energy spectra as pseudo-data, they consequently showed that these spectra could be
well reproduced within the much more restricted model space of the $1f_{7/2}$ orbital only, now containing
between 0 and 8 neutrons. Indeed, the effective interaction matrix elements adjusted to the spectrum
of the Ps nuclei corresponded to quite a different interaction than those in case of the larger model space.
Moreover, they showed that other observables, such as the $B(E2)$ values for the strongest transitions, 
were very similar in both approaches, even though the wave functions differed distinctly. 
A different set of Pseudonium nuclei was constructed for a model space of two degenerate 
$1p_{1/2}$ and $1d_{3/2}$  single-particle states that could contain both protons and neutrons,
up to a maximum of 12 nucleons. Very much the same
conclusion was reached after the analysis of the resulting spectra
within a restricted model space of the $1d_{3/2}$  
orbital only \cite{law67}. In the latter study, it was pointed out that quadrupole moments seemed 
to be a better observable to probe differences. Certain particularly chosen transfer reactions were
highly sensitive to the choice of the model space. This demonstrates that a number of observables
such as excitation energies and $B(E2)$-values are rather insensitive to configuration mixing arising from 
the excitation of zero-coupled pairs across the closed shell. The same underlying mechanism may
be responsible for the similarities between the results for the Pt-isotopes obtained within the $[N]$ 
configuration space of the IBM and those obtained for the $[N]\oplus[N+2]$ configuration space.
In addition to the detailed comparison in \cite{Garc09}, we have
constructed pseudo spectra in the IBM within a
$[N]\oplus[N+2]$ configuration space and consequently adjusted the parameters of an IBM 
Hamiltonian within the $[N]$ configuration space \cite{Garc-unp}. Apart from very particular
$B(E2)$ transition rates, it was impossible to discriminate between the results of the
two approaches.  

In a more recent example, a study of the actual wave function content
and the way to test it has been explored in the study of the nucleus
$^{40}$Ca \cite{caurier07}. It turns out that the $0^+$ ground state
consists of only 65\% closed sd shell (or 0p-0h) and exhibits 29\%
2p-2h excitations out of the $2s_{1/2},1d_{3/2}$ normally filled
orbitals into the $1f_{7/2}$, $2p_{3/2}$, $1f_{5/2}$, $2p_{1/2}$
higher-lying orbitals with even a 5\% 4p-4h excitation
contribution. This large model space is needed to describe the
higher-lying strongly deformed bands and superdeformation as
experimentally observed in $^{40}$Ca.  The isotopic shifts in the
even-even $A=40$ to $A=48$ Ca nuclei could be reproduced well through
explicit inclusion of mp-nh excitations across the $Z=20$,
$N=20$ "closed" shell in a slightly smaller model space than the one
mentioned before \cite{caurier01}. This indicates that one can
indeed find observables which are sensitive to the important
components of the wave function and thus can discriminate between
various approaches that give quite similar results when restricting to
a subset of data only.

The content of this paper is organised as follows. After 
Section \ref{sec-intro}, we present the IBM formalism in
Section \ref{sec-form}, the evolution of the
character of low-lying states in \ref{sec-evolution}, the systematics
of the energy spectra in \ref{sec-spectra}, and  the decomposition of 
the $B(E2)$ values into regular and intruder contributions in \ref{sec-E2}.
In Section \ref{sec-effect} we explain the origin of the observed flat energy
tendencies around neutron mid-shell on the basis of the crossing of the
regular and intruder unperturbed $0^+$ states. Section
\ref{sec-sensitiv} is devoted to the study of observables sensitive to
the presence of 2p-2h configuration such as gyromagnetic factors,
\ref{sec-gfactor}, $\alpha$-decay hindrance factors,
\ref{sec-alpha}, and nuclear radii, \ref{sec-radii}. Finally, in 
Section \ref{sec-conclu} we present the summary and the conclusions of
this work.

\section{Configuration mixing and the observation of regular patterns}
\label{sec-config-mix}
\subsection{The formalism}
\label{sec-form}
In this section, we present an abridged introduction to the IBM configuration
mixing formalism (or IBM-CM) and to the fitting-procedure of the IBM-CM
parameters for the Pt isotopes. For an in-depth discussion, we refer to \cite{Garc09}.
The IBM-CM allows the simultaneous treatment and mixing of several
boson configurations which correspond to different particle--hole
(p--h) shell-model excitations \cite{duval82}. On the basis of
intruder spin symmetry \cite{hey92,coster96}, no distinction is made
between particle- and hole-bosons. Hence, the model space including the valence
neutrons outside the $N=82$ shell as well as the regular 4h and 6h-2p proton 
configurations with respect to the $Z=82$ shell corresponds to a $[N]\oplus[N+2]$ 
boson space. The boson number $N$ is obtained as the sum of the number of active protons 
(counting both proton particles and holes) and the number of valence neutrons, divided by two.
Thus, the Hamiltonian for
two configuration mixing is written
\begin{equation}
  \hat{H}=\hat{P}^{\dag}_{N}\hat{H}^N_{\rm ecqf}\hat{P}_{N}+
  \hat{P}^{\dag}_{N+2}\left(\hat{H}^{N+2}_{\rm ecqf}+
    \Delta^{N+2}\right)\hat{P}_{N+2}\
  +\hat{V}_{\rm mix}^{N,N+2}~,
\label{eq:ibmhamiltonian}
\end{equation}
where $\hat{P}_{N}$ and $\hat{P}_{N+2}$ are projection operators onto
the $[N]$ and the $[N+2]$ boson spaces 
respectively, $\hat{V}_{\rm mix}^{N,N+2}$  describes
the mixing between the $[N]$ and the $[N+2]$ boson subspaces, and
\begin{equation}
  \hat{H}^i_{\rm ecqf}=\varepsilon_i \hat{n}_d+\kappa'_i
  \hat{L}\cdot\hat{L}+
  \kappa_i
  \hat{Q}(\chi_i)\cdot\hat{Q}(\chi_i), \label{eq:cqfhamiltonian}
\end{equation}
is the extended consistent-Q Hamiltonian (ECQF) \cite{warner83} with $i=N,N+2$,
$\hat{n}_d$ the $d$ boson number operator, 
\begin{equation}
  \hat{L}_\mu=[d^\dag\times\tilde{d}]^{(1)}_\mu ,
\label{eq:loperator}
\end{equation}
the angular momentum operator, and
\begin{equation}
  \hat{Q}_\mu(\chi_i)=[s^\dag\times\tilde{d}+ d^\dag\times
  s]^{(2)}_\mu+\chi_i[d^\dag\times\tilde{d}]^{(2)}_\mu~,\label{eq:quadrupoleop}
\end{equation}
the quadrupole operator. 
We did not consider the most general
IBM Hamiltonian for each Hilbert space, $[N]$ and $[N+2]$, but restricted ourselves to the
ECQF formalism Hamiltonian \cite{warner83,lipas85}. This 
approach has been shown to be a rather good approximation in many
calculations. 

The parameter $\Delta^{N+2}$ can be
associated with the energy needed to excite two particles across the
$Z=82$ shell gap, corrected for the pairing interaction energy gain and including
monopole effects~\cite{hey85,hey87}.
The operator $\hat{V}_{\rm mix}^{N,N+2}$ describes the mixing between
the $N$ and the $N+2$ configurations and is defined as
\begin{equation}
  \hat{V}_{\rm mix}^{N,N+2}=w_0^{N,N+2}(s^\dag\times s^\dag + s\times
  s)+w_2^{N,N+2} (d^\dag\times d^\dag+\tilde{d}\times \tilde{d})^{(0)}.
\label{eq:vmix}
\end{equation}

The $E2$ transition operator for two-configuration mixing is
subsequently defined as
\begin{equation}
  \hat{T}(E2)_\mu=\sum_{i=N,N+2} e_i
  \hat{P}_i^\dag\hat{Q}_\mu(\chi_i)\hat{P}_i~,\label{eq:e2operator}
\end{equation}
where the $e_i$ ($i=N,N+2$) are the effective boson charges and
$\hat{Q}_\mu(\chi_i)$ is the quadrupole operator defined in equation
(\ref{eq:quadrupoleop}).

In our fitting procedure, we focussed on obtaining the best
possible agreement with the experimental data available for the excitation energies
and for the $B(E2)$ reduced transition probabilities. In the most general case
13 parameters need to be determined for the IBM-CM Hamiltonian (\ref{eq:ibmhamiltonian})
and the $E2$ operator (\ref{eq:e2operator}).
To obtain parameters that vary smoothly from isotope to isotope, we imposed some constraints. 
For the regular Hamiltonian, we chose $\chi_N=0$, while
we fixed the relative d-boson energy to
$\varepsilon_{N+2}=0$ for the intruder Hamiltonian, the latter choice also supported by \cite{harder97}. 
These choices were made following a number of test calculations
in which no substantial improvement in the value of $\chi^2$ 
was obtained if we allowed $\varepsilon_{N+2}\neq 0$ or $\chi_N \neq 0$.
In addition, we kept the value that describes the
energy needed to create an extra particle-hole pair (or $2$ extra bosons) at
$\Delta^{N+2}=2800$ keV (note the typo $\Delta^{N+2}$=1400 keV in \cite{Garc09}; all calculations
were performed with the correct value, though). Similarly, the
mixing strengths were chosen $w_0^{N,N+2}=w_2^{N,N+2}=50$ keV for all the Pt isotopes. 
Those values are known to be quite appropriate in this part 
of the nuclear chart \cite{harder97,king98}, although the choice of 
the mixing strength remains somewhat arbitrary \cite{harder97}. 
Finally, we also have to determine the effective charges of the $E2$ operator for each isotope.
With these choices, the number of parameters still to be determined for each nucleus is 8.

The parameters for the IBM-CM Hamiltonian resulting from the fitting procedure
are summarised in Table \ref{tab-fit-par-mix} \cite{Garc09}. 
Note that some of the Hamiltonian parameters,
especially for $^{172}$Pt and $^{174}$Pt, remain rather arbitrary due to
the lack of experimental data. 
For $^{172}$Pt and $^{174}$Pt,
the value of the effective charges cannot be determined
because not a single absolute $B(E2)$ value is known. 
Similarly, for $^{182}$Pt, the absolute value of the effective charges could not be
determined because only relative $B(E2)$ values are known. As a consequence, those
charges are dimensionless.
\begin{table}
\begin{center}
\begin{tabular}{|c||c|c|c||c|c|c||c|c|}
\hline
Nucleus&$\varepsilon_N$&$\kappa'_N$&$\kappa_N$&$\kappa'_{N+2}$&
$\kappa_{N+2}$&$\chi_{N+2}$&$e_{N}$&$e_{N+2}$\\
\hline
$^{172}$Pt&725.0&0.00  &-39.47&0.00 &-22.87&-0.38 &-&-\\
$^{174}$Pt&701.2&0.00  &-31.60&0.00 &-21.82&-0.30 &-&-\\  
$^{176}$Pt&683.4&1.04  &-37.62&5.24 &-23.56&-0.75 &1.86&1.63\\  
$^{178}$Pt&753.8&-2.31 &-37.45&5.27 &-25.17&-0.55 &3.21&1.52\\  
$^{180}$Pt&999.3&-15.14&-37.34&6.57 &-25.14&-0.32 &1.29 &1.94 \\  
$^{182}$Pt&939.9&-6.70 &-35.39&7.03 &-23.50&-0.31 &1&1.1\\  
$^{184}$Pt&750.6&1.47  &-32.66&6.64 &-23.89&-0.34 &1.14 &1.71 \\  
$^{186}$Pt&675.3&3.17  &-30.50&7.29 &-24.23&-0.32 &1.44 &1.67 \\  
$^{188}$Pt&483.2&4.94  &-37.38&6.67 &-31.47&-0.11 &1.66 &1.66 \\  
$^{190}$Pt&338.7&19.33 &-34.62&0.83 &-32.51&0.00  &1.50 &1.50 \\  
$^{192}$Pt&314.9&12.01 &-45.32&-8.82&-38.84&0.00  &1.68&1.77\\  
$^{194}$Pt&370.9&6.67  &-38.26&6.52 &-31.02&0.00  &1.97&0.25\\  
\hline
\end{tabular}
\end{center}
\caption{Hamiltonian and $\hat{T}(E2)$ parameters resulting from the present study.
All quantities have the dimension of energy (given in units of keV),
except $\chi_{N+2}$ which is dimensionless and $e_{N}$ and $e_{N+2}$
which are given in units $\sqrt{\mbox{W.u.}}$ 
The remaining parameters of the Hamiltonian, i.e.~$\chi_N$ and $\varepsilon_{N+2}$ are equal to
zero, except  $\Delta^{N+2}$=2800 keV  and $w_0^{N,N+2}=w_2^{N,N+2}$=50 keV.}
\label{tab-fit-par-mix}
\end{table}

\subsection{The evolution of the character of the yrast band}
\label{sec-evolution}

We start our analysis with the structure of the configuration mixed wave functions along the yrast levels, expressed
as a function of the $[N]$ and $[N+2]$ basis states, 
\begin{eqnarray}
\Psi(k,JM) &=& \sum_{i} a^{k}_i(J;N) \psi((sd)^{N}_{i};JM) 
\nonumber\\
&+& \
\sum_{j} b^{k}_j(J;N+2)\psi((sd)^{N+2}_{j};JM)~,
\label{eq:wf:U5}
\end{eqnarray}
where $k$, $i$, and $j$ are rank numbers.

\begin{figure}[hbt]
  \centering
  \epsfig{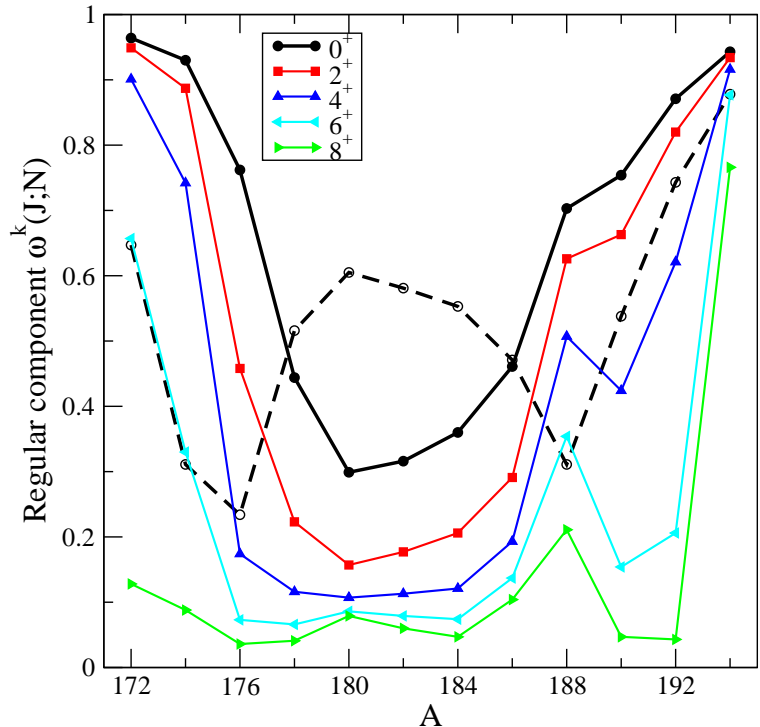} 
  \caption{(Color online)  Regular component of the yrast band states (full lines), together with
    the 0$^+_2$ state (black dashed line), calculated using the IBM-CM formalism.}
  \label{fig-composition}
\end{figure}
\begin{figure}[hbt]
  \centering
  \epsfig{file=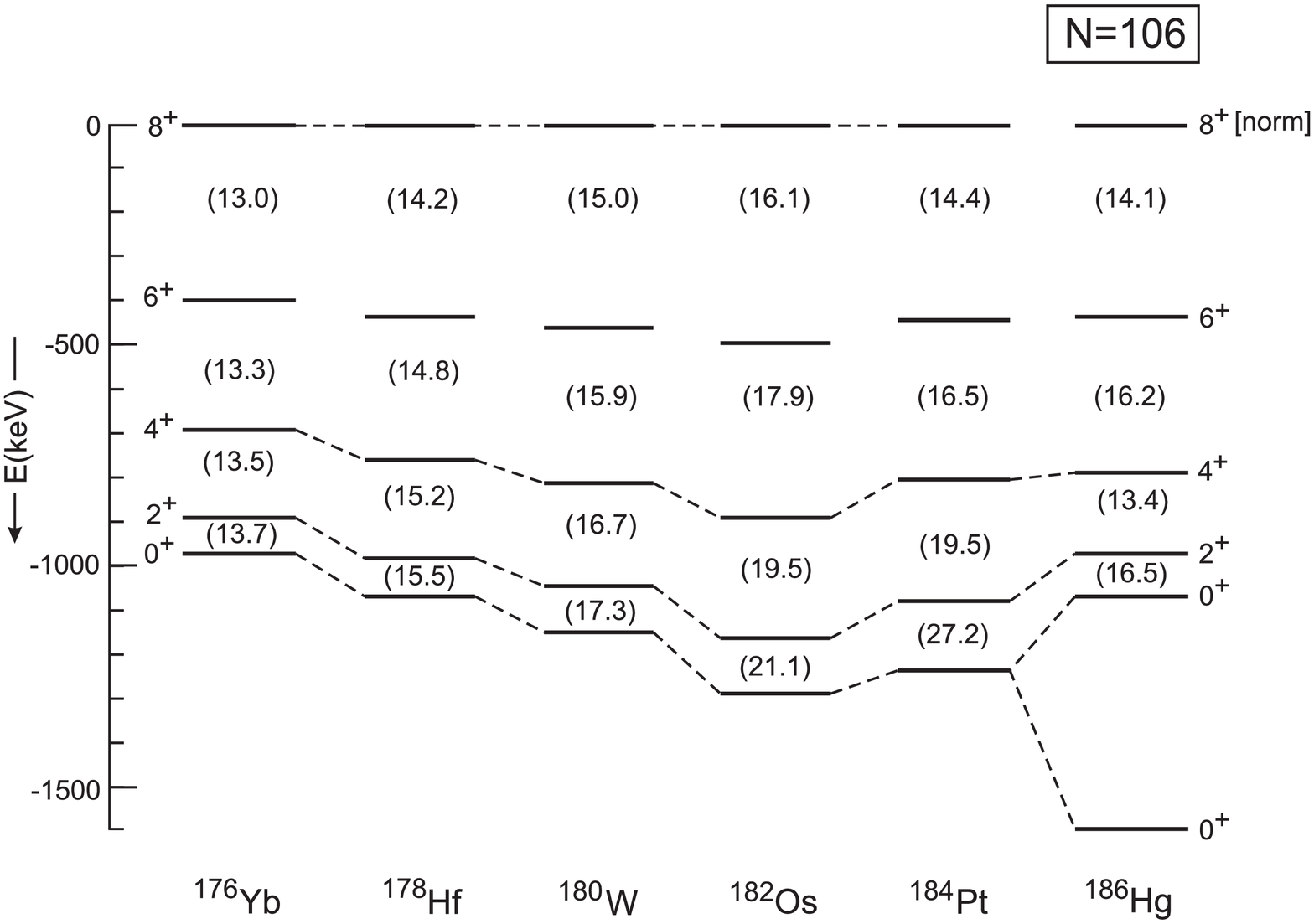,width=12cm} 
  \caption{Energy spectra in the $N=106$ isotones from  $_{70}$Yb up
    to $_{80}$Hg,  relative to the energy of the yrast 8$^+$ state. The
numbers between brackets denote the value of $\hbar^2/(2\cal{J})$
deduced from the energy differences.}  
  \label{fig-relative}
\end{figure}

In Fig.~\ref{fig-composition}, we present 
the weight of the wave function contained within the $[N]$-boson subspace, defined as
the sum of the squared amplitudes $w^k(J,N) \equiv \sum_{i}\mid a^{k}_i(J;N)\mid ^2$, for
the yrast states $(k=1)$ and the 0$^+_2$ state, which
is indicated with a dashed line.  
The results exhibit an interesting behaviour, both as a function of
angular momentum $J$ and as a function of  
changing mass number. Indeed, the character of the yrast band changes 
with increasing neutron number, passing from
being spherical (major component in the $[N]$ space) at mass number A
$\sim$172 towards more deformed  
(major component in the $[N+2]$ model space) close to mid-shell, and
changing again to a spherical character when  
approaching A$\sim$192. This behaviour is very pronounced for the yrast
0$^+_1$, 2$^+_1$, 4$^+_1$ states but changes for the 
higher spin states and in particular for the $J=8^+$ state, which retains its $[N+2]$ intruder
character along the whole region $172 \leq A \leq 192$ (except  for $^{194}$Pt which
is regular). This makes the $J=8^+$ state the ideal reference state to redraw
the energy spectra of the Pt nuclei and study their evolution \cite{wood81,wood82}. 
Similarly scaled energy spectra can be obtained for other nuclei exhibiting this systematic behaviour.
Hence, rescaling the energy spectra of the adjacent isotones with
neutron number $N=106$ (with $w^1(J=0,N) \sim 30\%$ for $^{184}$Pt)
gives a most interesting illustration that reveals the mixing effects in
the ground-state and lower-spin yrast states (see
Fig.~\ref{fig-relative}). It shows that the energy of the 0$^+$
ground state and some of the lower-spin yrast states relative to a
higher-lying, more pure, reference state is particularly sensitive to
the $[N+2]$ configuration space wave function components.  This is
studied in more detail in Sec.~\ref{sec-sensitiv}.  It is also clear
that the yrast band in $^{184}$Pt shows a very strong correspondence
with the intruder band structure in $^{186}$Hg.

Going back to Fig.~\ref{fig-composition}, where we also plotted the regular component of
the $0_{2}^{+}$ state, one clearly notices its complementary behaviour compared to the $0_1^+$ state. This has an important 
consequence on the study of the hindrance factor for $\alpha$ decay from the Hg ground state into the 
0$^+_{1,2}$ states in the Pt nuclei, as will be discussed in Section \ref{sec-alpha}.


\subsection{Energy spectra}
\label{sec-spectra}

Having discussed the wave function content
in terms of the $[N]$ and $[N+2]$ configurations in the previous section,
as a next step we study the configuration mixed energy spectra
in more detail. Especially the energies up to E$_x \sim$ 1.5 MeV
are of interest because the extra states coming from the intruder
configuration do not show up in an obvious way (in contrast to, e.g.
the  Pb and Hg nuclei in Fig.~\ref{fig-Pb-syst} and
Fig.~\ref{fig-Hg-syst}, respectively). Therefore, we diagonalize the Hamiltonian
(\ref{eq:ibmhamiltonian}) without the mixing term in the separate
model spaces $[N]$ and $[N+2]$. This results in the wave functions 
\begin{equation}
\Psi(k,JM)_N = \sum_{l} c^{k}_{l}(J;N) \psi((sd)^{N}_{l};JM) ~,
\label{eq:wf:unpert}
\end{equation}
(and similarly for $N+2$) with corresponding energies E(k,J[N]) and E(l,J[N+2]).
This method has been described in detail \cite{fossion03,helle05,helle08} and results in ``so-called''
unperturbed bands that are an intermediate step before obtaining the full, configuration-mixed,
wave functions and their corresponding excitation energies.
These bands are the equivalent of the unperturbed bands extracted from phenomenological band mixing studies that
have been carried out in this mass region \cite{dracou04,page03}.
The introduction of the coupling term $\hat{V}_{\rm mix}^{N,N+2}$ leads to a mixing of these
unperturbed bands. The intermediate basis of a set of unperturbed
``bands'' is particularly useful to detect the effects caused by 
the remaining mixing term and its influence on the final energy spectra.

\begin{figure}[hbt]
  \centering
  \epsfig{file=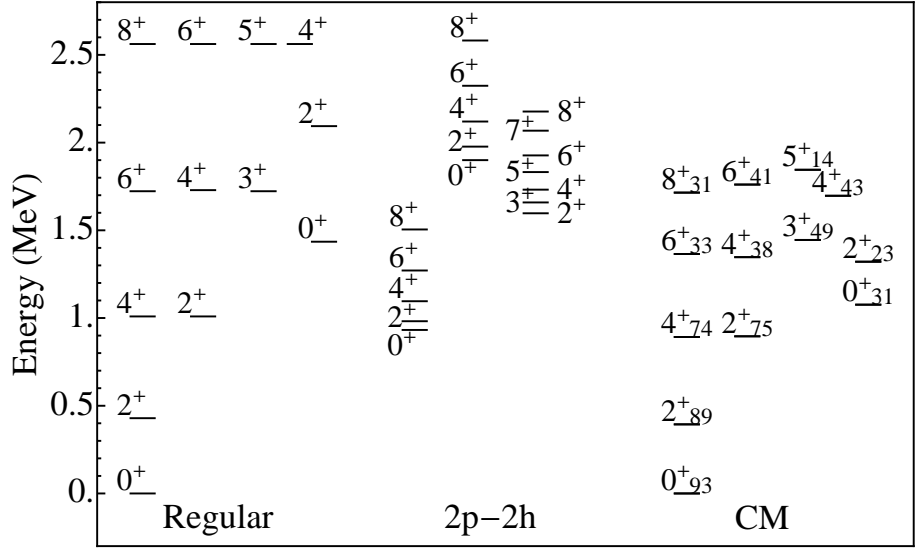,width=12cm} 
  \caption{Unperturbed regular and intruder (``2p-2h'') energy levels together
    with the theoretical fully mixed calculation (``CM'') for
    $^{174}$Pt. The small numbers in the ``CM'' column correspond to the
    regular component percentage.}
  \label{fig-174pt-ener}
\end{figure}

In the next figures, we illustrate these various steps highlighting
the way in which the energy spectra result. We consider, as examples,
the nuclei $^{174}$Pt, $^{180}$Pt, $^{186}$Pt, and $^{192}$Pt which are
positioned around neutron mid-shell $N=104$. Starting with $^{174}$Pt ($N=96$),
we present the unperturbed bands resulting from diagonalizing in the
$[N]$ space (called ``Regular'') and in the $[N+2]$ space (called
``2p-2h'') on the left-hand side of Fig.~\ref{fig-174pt-ener}. The lowest
unperturbed regular bands correspond to the less collective
structure whereas the higher-lying unperturbed 2p-2h bands have
a rotational-like structure, including bands that resemble excited
K$^{\pi}$=2$^+$ and K$^{\pi}$=0$^+$ bands. The inclusion of the mixing
then leads to the energy spectra (called CM for ``configuration
mixed'') at the right-hand side of the figure. On each of the levels,
the weight $w^k(J,N) \equiv \sum_{i}\mid a^{k}_i(J;N)\mid ^2$ (see
Section \ref{sec-evolution} for its definition) of the regular $[N]$
part of the model space is displayed. This nicely illustrates the
gradual degrading of the $[N]$ percentage when going up the yrast band
(see also Fig.~\ref{fig-composition}).  It also explains why the 2p-2h
intruder band is not clearly separated from the regular band
structure: the mixing induces a particular redistribution of the
energy levels such that lowest band members originating from the
unperturbed 2p-2h configuration end up in different final sets of
states.  We stress that the bands were constructed as sequences
of levels connected through large $B(E2)$ reduced transition
probabilities.  We also mention that only the lowest 3 bands are shown
in the CM spectrum as we focus on those bands which appear mostly
below $\sim$ 1.5 MeV.
Finally, it is clear that the CM spectrum strongly resembles that of
the regular configuration even though the energies are  
more compressed. Evidently, the wave functions are largely different in both cases.

\begin{figure}[hbt]
  \centering
  \epsfig{file=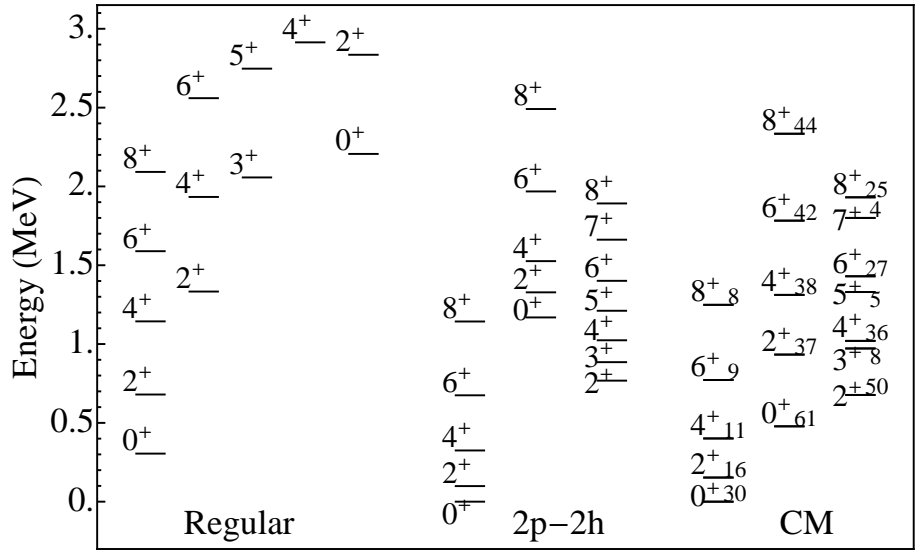,width=12cm} 
  \caption{The same as Fig.~\ref{fig-174pt-ener} but for $^{180}$Pt.} 
  \label{fig-180pt-ener}
\end{figure}

\begin{figure}[hbt]
  \centering
  \epsfig{file=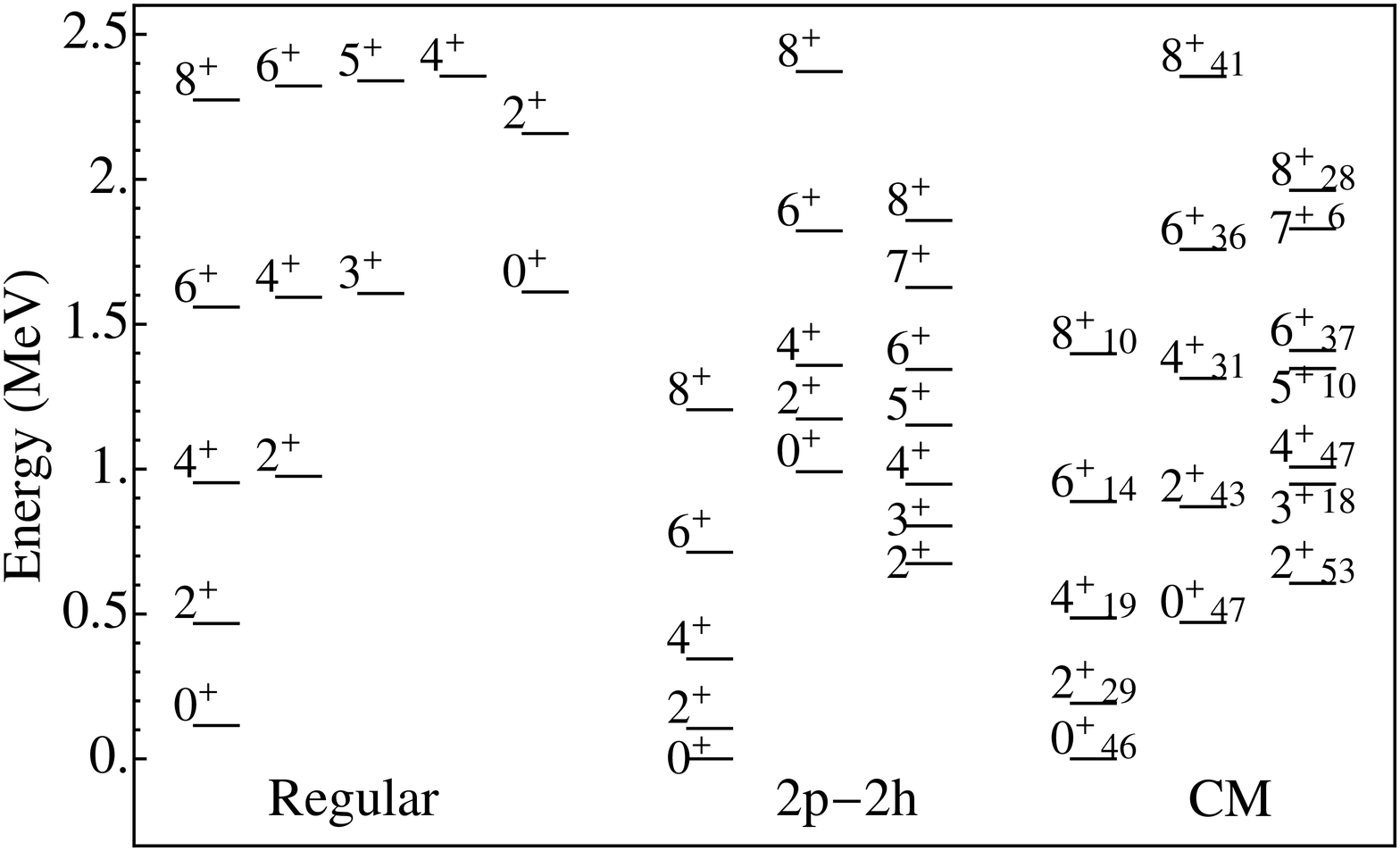,width=12cm} 
  \caption{The same as Fig.~\ref{fig-174pt-ener} but for $^{186}$Pt.}
  \label{fig-186pt-ener}
\end{figure}

In Fig.~\ref{fig-180pt-ener}, we illustrate the situation for
$^{180}$Pt ($N=102$), which is situated close to neutron mid-shell 
$N=104$. Here, one clearly notices two things as compared with
$^{174}$Pt: (i) the inversion of the energies of lowest 2p-2p 
unperturbed bands with the regular unperturbed bands, and (ii)
the change in structure of the regular unperturbed bands. 
When comparing with the CM spectrum, one notices that the yrast band
has its main components within the $[N+2]$ model space,  
becoming gradually pure with increasing angular momentum. 
The even angular momentum states in the two excited bands in the CM
spectrum retain mostly a $[N+2]$ character but with a  
much larger contribution from the $[N]$ components as compared to the
yrast band. In fact, they result mostly from strong  
mixing between the lowest unperturbed regular band and the two excited
unperturbed 2p-2h bands. 
In $^{186}$Pt ($N=108$, see Fig.~\ref{fig-186pt-ener}), the unperturbed bands are almost
degenerate. This is reflected in the composition of the wave
functions, in particular for the lower spin states where strongly 
mixed configurations result. 
For both $^{180}$Pt and $^{186}$Pt, the 
strong mixing of the unperturbed bands conceals
the presence of the two different configurations for the
bands in the CM spectrum starting off well below 1.5 MeV.

\begin{figure}[hbt]
  \centering
  \epsfig{file=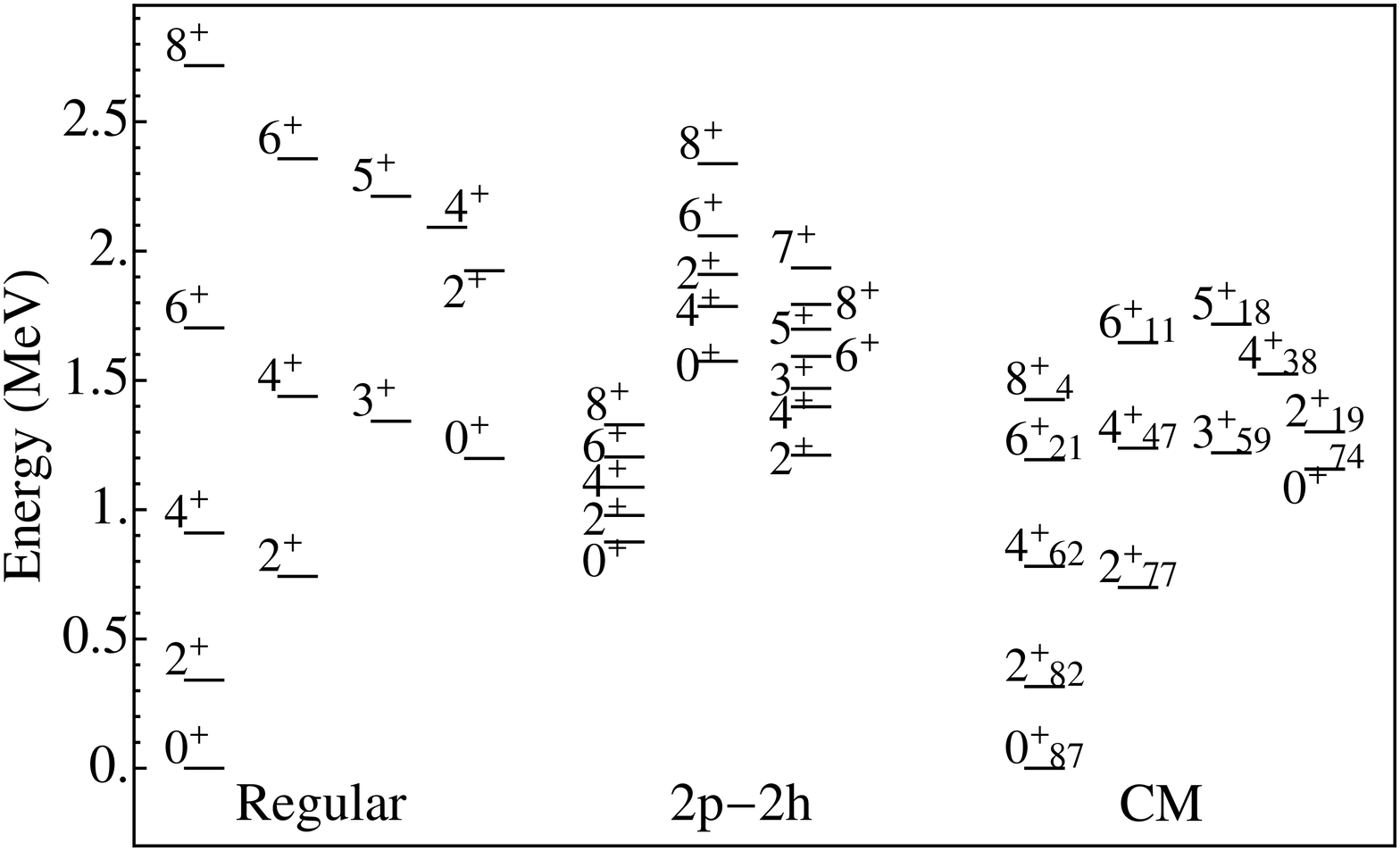,width=12cm} 
  \caption{The same as Fig.~\ref{fig-174pt-ener} but for $^{192}$Pt.}
  \label{fig-192pt-ener}
\end{figure}

For comparison, we also present in Fig.~\ref{fig-192pt-ener})the
results for $^{192}$Pt ($N=114$), a nucleus well past the 
neutron mid-shell $N=104$.  In this case, the
2p-2h unperturbed bands have moved up considerably compared to
the regular bands which exhibit a clear O(6) structure in this mass region. 
Even though the unperturbed regular and 2p-2h bands seem well separated at first sight, the 
states above the $2_{1}^{+}$ level remain quite mixed in the spectrum resulting after
configuration mixing.

As a conclusion to this part of our study, in which we investigated the unperturbed
bands (regular and 2p-2h) and subsequently added the mixing interaction, we state
that a very strong mixing of the bands for the Pt nuclei close to neutron mid-shell
makes it virtually impossible to distinguish between the regular and 2p-2h configurations. 
Because of the strong mixing in the Pt nuclei, in particular in the
mid-shell region, (i) it is hard to distinguish
the precise nature of a band by just observing the energy systematics,
and, (ii) remarkably, the energy spectra
resulting from the IBM-CM calculation, below $\sim$ 1.5 MeV, resemble spectra that can be described with an IBM Hamiltonian in the $[N]$ space.

\subsection{Electric quadrupole transitions}
\label{sec-E2}

More detailed information on the admixture of the wave functions can
be obtained from the $E2$ transition matrix 
elements. Whereas the wave functions in eq.~(\ref{eq:wf:U5}) are
expressed using the $[N]$ and $[N+2]$ basis states, 
we can equally well express them using the eigenfunctions
corresponding with the unperturbed regular and 
intruder bands, as given in eq.~(\ref{eq:wf:unpert}). 
Under this basis transformation, the $E2$ transition matrix elements are decomposed into
corresponding  $E2$ transition matrix elements within the unperturbed
bands each having a certain weight factor. 
This allows to filter out those transitions in the unperturbed bands
that contribute most to a certain transition 
in the fully configuration mixed bands and provides additional information on the admixture in 
the wave function.

\begin{figure}[hbt]
  \centering
  \epsfig{file=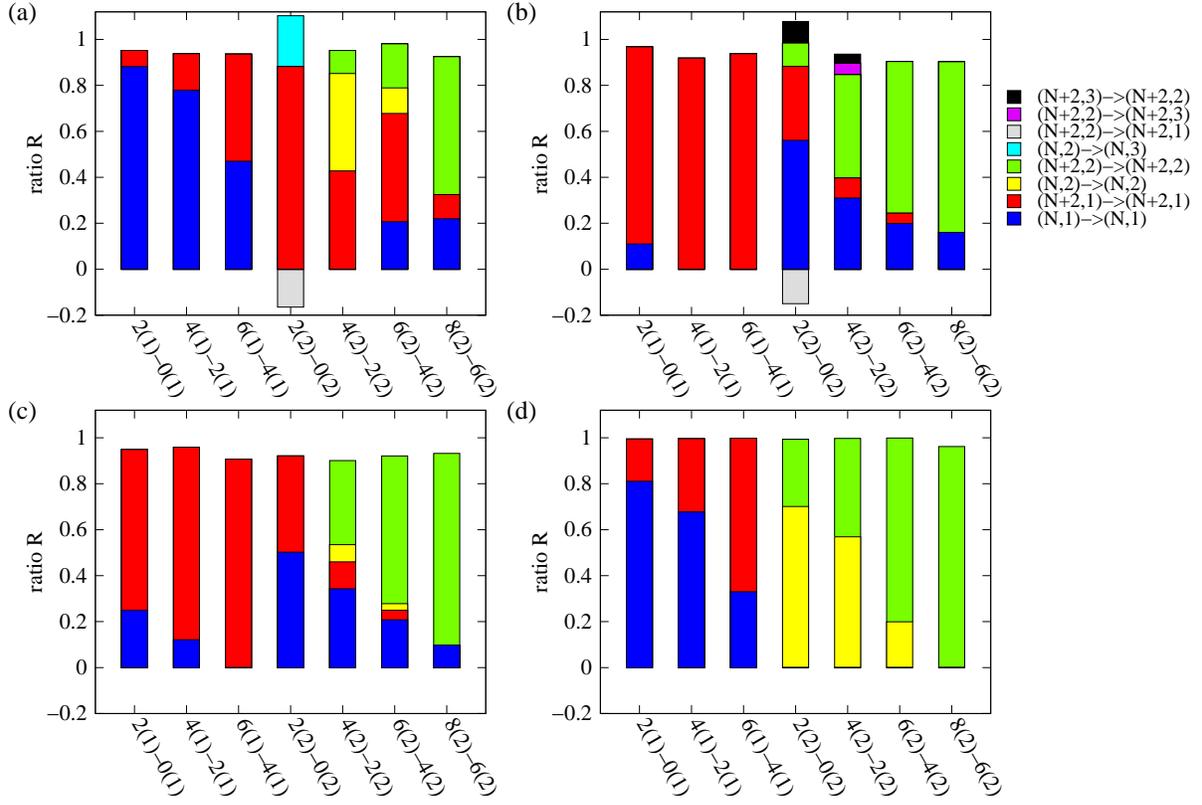,width=16cm} 
  \caption{(Color online) Decomposition of the $E2$ matrix elements for $^{174}$Pt
    (panel a), $^{180}$Pt
    (panel b), $^{186}$Pt
    (panel c), and $^{192}$Pt
    (panel d). The $E2$ ratio R is defined as in \cite{helle08}.
    The initial and final state are indicated as J$_i$(i)-J$_f$(f) at the bottom of the figure. The color coding
    for the contributions of the corresponding transition from  $J_{i}(i')$ to $J_{f}(f')$ in the unperturbed 
    regular band (indicated as (N,i')$\rightarrow$ (N,f')) is given at the right-hand side of the figure
     (similar for contributions in the unperturbed 2p-2h band).}
  \label{fig-180pt-E2}
\end{figure}

In Fig. \ref{fig-180pt-E2}, we present the ratio $R$ for (a) $^{174}$Pt,  (b) $^{180}$Pt, (c) $^{186}$Pt,
and (d) $^{192}$Pt. This quantity $R$ is defined as the ratio of a
contributing 
reduced transition matrix  
element of the $J_{i}(i')\rightarrow J_{f}(f')$ transition in the unperturbed regular band 
$\langle (f',J_{f})_{N}\mid\mid\hat{T}(E2)\mid\mid (i',J_{i})_{N}\rangle$  
(and similar for N+2) times its weight factor, with respect to the reduced transition matrix element  of the 
corresponding transition in the fully configuration-mixed system
$\langle (f,J_{f})\mid\mid\hat{T}(E2)\mid\mid
  (i,J_{i})\rangle$ it contributes to ($f',i',f,i$ being rank numbers).
We have plotted the most important contributions (ratios), such that, when adding them, we arrive to within
10\% of the full matrix element. 
The inset legend in Fig.~\ref{fig-180pt-E2} gives the specific contributions in the unperturbed band. For example, 
in blue $(N,1)\rightarrow (N,1)$, the contributing ratio 
\begin{equation}
R(N)=W~\frac{\langle (1, J-2)_{N}\mid\mid\hat{T}(E2)\mid\mid (1, J)_{N}\rangle}{\langle 
(k, J-2)\mid\mid\hat{T}(E2)\mid\mid (k, J)\rangle}~,
\end{equation}
with $W$ the weight factor (see \cite{helle05,helle08} for the
detailed expression), is shown (the same can be defined for $[N+2]$). 
The effective charges used for these decompositions are taken from
Table \ref{tab-fit-par-mix}, except for $^{174}$Pt where
arbitrary charges $e_{N}=e_{N+2}=1$ have been used.  

Inspecting the transitions in the yrast band (i.e. $2(1)-0(1)$,
$4(1)-2(1)$, and $6(1)-4(1)$), we observe a pattern to be expected
from the discussion of the energy spectra in
Sect. \ref{sec-spectra}. Whereas, the largest contributions are coming
from the unperturbed regular yrast band for $^{174}$Pt and $^{192}$Pt,
with an increasing contribution from the unperturbed intruder yrast
band when going to higher spin, the transitions in the yrast band of
$^{180}$Pt and $^{186}$Pt are almost entirely determined by the
contribution of transition in the unperturbed 2p-2h yrast band.

The transitions in the first excited band are more
interesting. Starting with $^{192}$Pt, we notice that those
transitions are dominated by the corresponding transitions in the
first excited unperturbed regular and 2p-2h band. Though this may seem
surprising at first sight from inspection of the spectrum, one should
keep in mind that the intruder part of the Hamiltonian is very close
to the case of $O(6)$ symmetry and that selection rules for the
transitions will apply. For the other nuclei, the decomposition of the
$E2$ matrix elements looks more complicated. For $^{180}$Pt and
$^{186}$Pt (excluding the $2(2)-0(2)$ transition for the moment), the
relatively largest contribution is coming from the transitions in the
first excited unperturbed intruder band, followed by contributions
from transitions in the unperturbed yrast regular band and some
smaller contributions. Indeed, the strong lowering of the unperturbed
intruder bands around neutron mid-shell and the typical spreading
of the energies in the unperturbed bands brings the $4^{+}$ and
$6^{+}$ states of the unperturbed regular yrast band and of the
unperturbed first excited 2p-2h band pretty close in energy. Even
though one would also expect a non-negligible contribution from the
transitions involving the second excited unperturbed 2p-2h band from
comparison of the unperturbed energies, they do not or barely
contribute. In $^{174}$Pt, finally, the unperturbed 2p-2h bands have
moved up in energy again and the $E2$ transitions in the first excited
band of the CM spectrum contain contributions from both yrast and
first excited unperturbed regular and 2p-2h band, indicating a wider
'spreading' of the wave function.  Finally, note that the $2(2)-0(2)$
transition, being an inter-band transition, often has a structure that
slightly differs from the rest of the second excited band.  Hence, we
may state that the decomposition of the E2-transition matrix elements
gives some more insight into the precise 'spreading' of the
wave function in the basis of the unperturbed states. The relative
purity of the $E2$ ratios within the yrast band also hints at the possibility
to describe those transitions in a reduced [N] space. From inspection of
the more complex structure of the $E2$ ratios in the first excited band and
 especially for the interband transition, one would expect differences between
 calculations with different model spaces to show up.

\section{Effect of configuration mixing on the systematics of energy
  levels} 
\label{sec-effect}

In this section, we concentrate on how the strong mixing effects 
discussed in Sect.~ \ref{sec-evolution}-\ref{sec-E2} give rise to the
characteristic energy  systematics of the even-even Pt nuclei.

\begin{figure}[hbt]
  \centering
  \epsfig{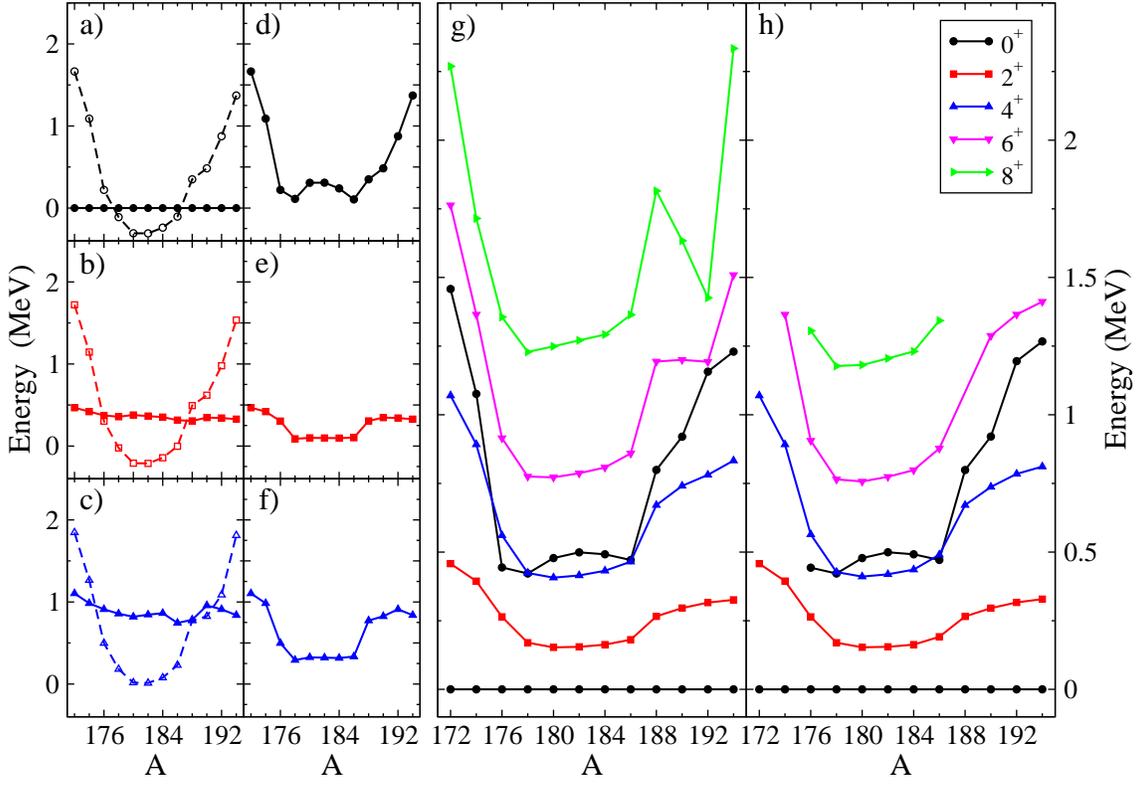} 
  \caption{(Color online) Energy systematic of the unperturbed (IBM-CM) first regular and
    intruder states (full line for regular, while dashed line for
    intruder states in panels a), b), and c)),
    a) $0_{1,2}^+$ energies relative to the energy of the $0^+$
    regular state,
    b) $2_{1,2}^+$ energies relative to the energy of the $0^+$
    regular state,
    c) $4_{1,2}^+$ energies relative to the energy of the $0^+$
    regular state,
    d) $0_{2}^+$ energies relative to the energy of the $0_1^+$
    state, 
    e) $2_{1}^+$ energies relative to the energy of the $0_1^+$
    state,
    f) $4_{1}^+$ energies relative to the energy of the $0_1^+$
    state. Panel g) corresponds to the full (including now all states)
    IBM-CM calculation and panel h) shows the experimental data of
    the yrast band states plus the $0_2^+$ state.}
  \label{fig-2level_energy}
\end{figure}

In panels a), b), and c) of Fig.~\ref{fig-2level_energy}, we plot the
energy systematics of respectively the unperturbed $0_{1,2}^+$, $2_{1,2}^+$,
and $4_{1,2}^+$ states. The energies are plotted relative to the energy 
of the unperturbed regular $0^+$ state, which enhances the parabolic behaviour of the energy 
intruder band. To compare those unperturbed energies to the final spectrum,
we should plot them relative to the energy of the $0^+$ state that is
lowest in energy. The unperturbed energies of the $0_{2}^+$, $2_{1}^+$,
and $4_{1}^+$ states with respect to the energy of the $0^{+}$ state are plotted
in panels d), e), and f), respectively. We
observe a very striking tendency due to the crossing of the $0_1^+$
and $0_2^+$ states. The evolution of $0_2^+$ state is mexican hat shaped
whereas the $2^+_1$ and $4^+_1$ states exhibit an almost flat behaviour around mid shell. 
This characteristic behaviour is exclusively due to
the crossing of $0_1^+$ and $0_2^+$ states. 
Upon inclusion of the mixing interaction, it is clear that the mixing effect will be maximal 
near the crossing at mass number $A=178$ and $A=186$,
in particular for the $0^{+}$ states and likewise for the higher angular momentum states. 
It is the interplay of the crossing with subsequent mixing for the 0$^+$ states
that largely determines the final behaviour of the energy systematics. 
The resulting spectra up to spin 8$^+$ are shown in
panel g) and still clearly display the very specific pattern of the 0$^+_2$ and 4$^+_1$ states
after the mixing. When comparing those same states with the experimental
systematics (see panel h), a clear-cut correspondence shows up.

Therefore, we can conclude that the crossing of the set of unperturbed regular and intruder bands 
is of major importance to describe the energy systematics as observed in
the Pt nuclei, even though the configuration mixing effects are highly
concealed in the energy levels and $B(E2)$ values of a given nucleus.

\section{Study of observables sensitive to configuration mixing}
\label{sec-sensitiv}

Following from our discussion in the previous sections, it should be
clear that nuclear structure effects caused by the strongly changing character 
of the wave function in the $[N]$ and $[N+2]$ space are to be expected for a number
of variables. Indeed, observables such as charge radii, gyromagnetic factors,
and $\alpha$-decay hindrance factors are sensitive to an increased number
of active protons (generated through particle-hole pair excitations across the
$Z=82$ closed shell), or to a change towards more explicit prolate
deformation near neutron mid-shell $(N=104)$. Therefore,
we will focus on these experimental quantities as they allow to probe
precisely those components of the nuclear wave functions.

\subsection{Gyromagnetic factors}
\label{sec-gfactor}

A, for our purpose, particularly interesting set of data are the g-factor
measurements for the $2^+_1$ state in the mid-shell $^{184,186,188}$Pt nuclei
\cite{stuchbery96}. The data display a rather flat behaviour 
as a function of the neutron
number in the vicinity of mid-shell. Early calculations by Kumar and Baranger
that were quite consistent with the data
\cite{kumar68} indicated a change from a prolate towards a more
oblate ground-state shape between $A=188$ and $A=190$ and were later
substantiated by studies from Bengtsson {\it et
  al.}~\cite{bengt87}. Stuchbery {\it et al.}~\cite{stuchbery96}
analysed gyromagnetic factors starting from the
two-band mixing study carried out by
Dracoulis {\it et al.}~\cite{dracou86}, in which the mixing between a
regular and an intruder configuration consistent with the measured
$B(E2)$ and with $E0$ measurements by Xu {\it
  et al.}~\cite{xu92} was extracted. 
The calculations by Stuchbery {\it et al.}~\cite{stuchbery96} 
pointed out that the
data cannot be described using only a single configuration but are
consistent with the mixing of two configurations. In particular,
the need of an increased number of active proton pairs for
the description of the $A=184, 186, 188$ results was demonstrated.  The same conclusion was
reached by Harder {\it et al.}~\cite{harder97}. More recently,
Bian {\it et al.}~\cite{bian07} carried out  projected
shell-model calculations starting from a deformed basis,
concentrating on g-factors for the $2^+_1$ state throughout the
rare-earth region, i.e. from Gd up to the Pt nuclei. Although they obtained a
rather good agreement for most of the region, the calculated results
show a distinct set of too low g-factors in the Pt nuclei in the mass
region $184 \leq A \leq 198$. Only by means of an artificial increase 
of the deformation, one could improve the agreement. Thus, g-factors
are sensitive observables to the precise configuration content
of the nuclear wave function describing the $2^+_1$ state.

\begin{figure}[hbt]
  \centering
  \epsfig{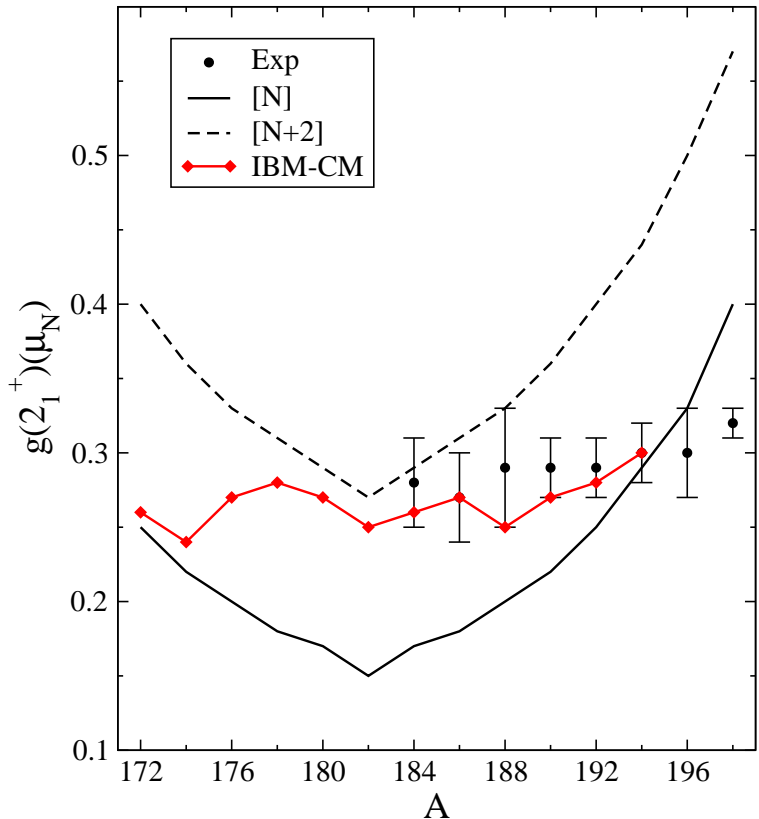} 
  \caption{(Color online) Gyromagnetic factor for the even-even Pt isotopes
    (experimental data from \cite{bian07}). Full circles for the
    experimental data, full and dashed lines for $[N]$ and $[N+2]$
    unperturbed results, respectively, and red full diamonds with full
    line for the IBM-CM calculations.} 
  \label{fig-g2_1}
\end{figure}

Within an IBM context, magnetic moments can be calculated with the IBM-2 \cite{Arim77,Otsu78},
which differentiates between proton ($\pi$) and neutron bosons
($\nu$) . The M1 operator can then be written as
\begin{equation}
\hat{T}(M1)=\sqrt{\frac{3}{4\pi}} \Big(\hat{P}^{\dag}_{N}(g_N^\pi
\hat{L}_\pi+g_N^\nu \hat{L}_\nu )\hat{P}_{N}+ 
\hat{P}^{\dag}_{N+2}(g_{N+2}^\pi
\hat{L}_\pi+g_{N+2}^\nu \hat{L}_\nu )\hat{P}_{N+2}\Big). 
\label{TM1}
\end{equation}
Using the standard microscopic values for the g factors \cite{Neumann95},
i.e.~$g_{N}^{\nu}=g_{N+2}^\nu=0$ and 
$g_{N}^{\pi}=g_{N+2}^\pi=\mu_N$, the M1 operator reduces to,
\begin{equation}
\hat{T}(M1)=\sqrt{\frac{3}{4\pi}} \Big(\hat{P}^{\dag}_{N}(\hat{L}_\pi)\hat{P}_{N}+ 
\hat{P}^{\dag}_{N+2}(\hat{L}_\pi)\hat{P}_{N+2}\Big)\mu_N. 
\label{TM1_red}
\end{equation}
The calculation of the matrix element of this operator cannot be
accomplished directly with IBM-1, but if one assumes F-spin symmetry for
the IBM-2 Hamiltonian \cite{Otsu78}, it can be readily shown that the gyromagnetic
factor can be written as \cite{harder97},
\begin{equation}
\frac{g(2_1^+)}{\mu_N }=\frac{1}{2 \mu_N } \mu(2_1^+)=\frac{N_\pi}{N}\omega^1(2,N)+\frac{N_\pi+2}{N+2}(1-\omega^1(2,N)),
\end{equation}
where $N_\pi$ is the number of protons out of the closed shell divided
by two and $\omega^1(2,N)$ is that part of the wave function of the 
$2_1^+$ state within the $[N]$-boson (regular) space (see Section
\ref{sec-evolution} for its definition). 
In Fig.~\ref{fig-g2_1}, we present
the calculated g-factors and the experimental values. 
This figure is qualitatively similar to the one in
\cite{harder97}, but displays a better agreement with the experimental data.
Note that this calculation is parameter free once 
the wave functions are determined.
As a reference, we also plotted the limits corresponding to wave
function with either fully regular $[N]$ character or intruder
$[N+2]$. The theoretical results obtained after the mixing calculation
should be situated between both lines.  Note that, according to
the IBM, this flat behaviour of the g-factors is necessarily explained by a strong
mixing between the regular and intruder 2p-2h configurations. 


\subsection{$\alpha$-decay hindrance factors}
\label{sec-alpha}

In the Pb-region, most interesting results were obtained when the
content of the nuclear wave functions was tested through
$\alpha$-decay measurements.  It was shown by Andreyev {\it et al.}~\cite{andrei00}
that $\alpha$-decay was instrumental as a sensitive probe to prove the presence
of a triplet of $0^+$ states in $^{186}$Pb, each corresponding to a different
shape.

Wauters {\it et al.}~\cite{wauters94,wauters94a} carried out experiments
on the $\alpha$-decay from the Po, Pb and Hg nuclei to the Pb, Hg and Pt nuclei,
respectively, concentrating in particular on the $N=104$ mid-shell
region. $\alpha$ decay is a highly sensitive fingerprint, precisely
because an $\alpha$ particle is emitted in the decay,  a process which requires
the extraction of two protons and two neutrons from the initial nucleus.  
The comparison of s-wave {\it l}=0 $\alpha$-decay branches from a given
parent nucleus (the Hg $0^+$ ground state in the present situation) 
to $0^+$ states in the daughter nucleus (the Pt $0^+$ ground state and
excited $0^+$ states) is important in that respect.
\begin{figure}[hbt]
  \centering
  \epsfig{file=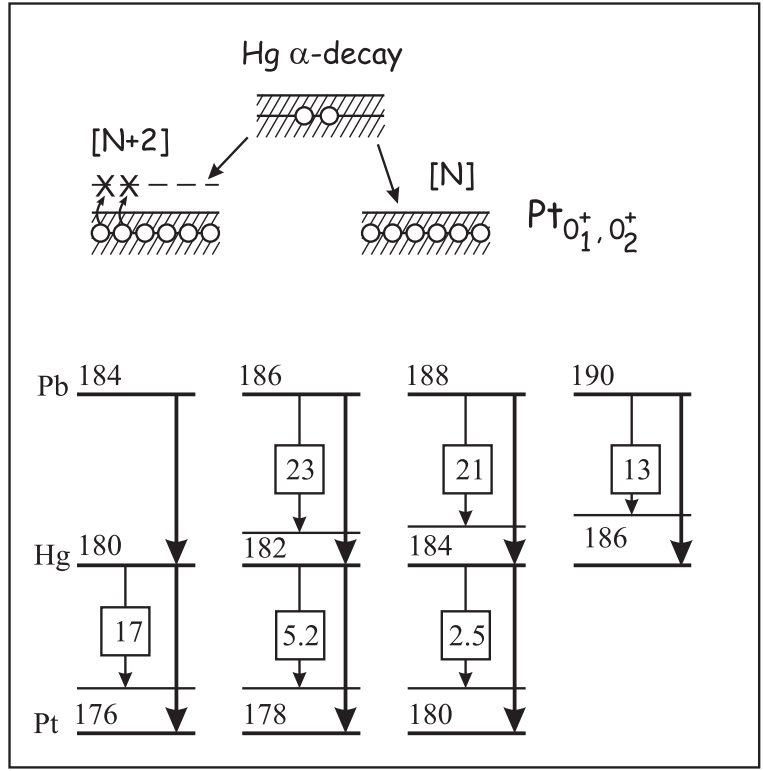,width=10cm} 
  \caption{A schematic view of the $\alpha$-decay proceeding from the 0$^+$ Hg ground
  state in the 0$^+_{1,2}$ states in the Pt nuclei. The inset boxes present the experimental $\alpha$ decay hindrance
  factors, which were taken from 
  \cite{wauters94,wauters94a,duppen00} and from Nucl. Data Sheets.}
  \label{fig-alpha}
\end{figure}
The reduced $\alpha$-decay widths themselves are very difficult to 
calculate on an absolute scale, but hindrance factors reflect possible changes amongst the wave functions
describing various $0^+$ states in a given daughter nucleus \cite{duppen00} well.
Hindrance factors of an $\alpha$-decay branch to an excited state with a strength $I_{ex}$,
relative to the $\alpha$-decay branch to the ground state with intensity $I_{gs}$ are defined by the
ratio
\begin{equation}
HF = \frac{\delta_{gs}^2}{\delta_{ex}^2}=\frac{I_{gs}P_{ex}}{I_{ex}P_{gs}} ,
\end{equation} 
where $\delta_{i}^2$ is the reduced $\alpha$ width, $P_{\alpha_i}$ the
penetration probability through the combined Coulomb 
and centrifugal barrier \cite{duppen00} and $I_i$ the $\alpha$-decay intensity
(with $i= gs, ex$ for the ground state and excited state, respectively).
They indicated that,
in the neutron mid-shell region, the $0^+$ ground-state in the Pb and Hg
nuclei is essentially consistent with a closed $Z=82$ core and a
two-proton hole configuration in the $Z=82$ core \cite{wauters94,wauters94a}
(see upper part of Fig.~\ref{fig-alpha}; only the proton structure is depicted,
as one does not expect the neutron part to be different in the final states). 
However, $\alpha$-decay feeding into the   
first-excited $0^+_2$ state exhibits a hindrance factor 
which is increasing with decreasing mass number
(see Fig.~\ref{fig-alpha}, lower part. The specific values of the hindrance factors are the adopted values as given in Nucl.
Data Sheets, starting from the original data \cite{wauters94,wauters94a}).
The observed large increase in hindrance when moving towards $N=100$
($A=178$) is consistent with the two-band mixing calculations by
Dracoulis {\it et al.}~\cite{dracou86}  which results in a 0$^+$ ground state exhibiting an increasing regular $[N]$ configuration
weight of $\approx$ 50\% for mass A=180 and A=178 up to $\approx$ 80\% for mass A=176.   
This is consistent with the results presented
in Fig.~\ref{fig-composition}, where the the $0^+_1$ ground state is progressively becoming 
a regular $[N]$ configuration, moving from mass $A=180$ (with $\approx 30 \%$ of
$[N]$ component) towards $A=178$ ($\approx 45 \%$ of
$[N]$ component)  and $A=176$ ($\approx 75 \%$ of $[N]$ component).
The important point here, as also stressed by Van Duppen and Huyse \cite{duppen00}, 
is the consistent picture that results when
treating the Po, Pb, Hg, and Pt nuclei jointly. More detailed calculations have been carried out
by Delion {\it et al.}~\cite{delion95,delion96}, and more recently by
Karlgren {\it et al.}~\cite{karlgren06}, 
and by Xu {\it et al.}~\cite{xu07},
emphasising the need for a microscopic QRPA description that encompasses both neutron and 
proton pairing vibrations and that includes proton 2p-2h excitations across the $Z=82$ closed shell.
They calculated hindrance factors for $\alpha$-decay into the neutron-deficient 
Po, Pb, Hg, and Pt nuclei. The hindrance factors for decay into the $^{176,178,180}$Pt
first excited $0^+_2$ state exhibit a large increase 
when moving down from mass $A=180$ towards $A=176$, corroborating
the results from a simple two-level analysis \cite{wauters94a}.  
Thus, $\alpha$-decay hindrance factors can serve as a 
sensitive fingerprint  to test structural changes of the nuclear wave functions.

\subsection{Isotopic shifts}
\label{sec-radii}

\begin{figure}[hbt]
  \centering
  \epsfig{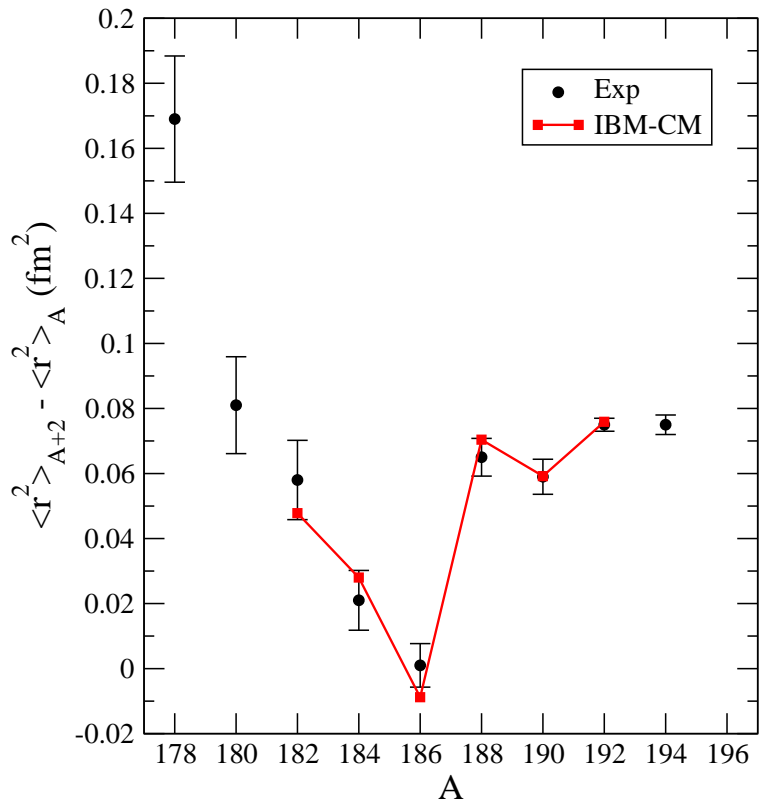} 
  \caption{(Color online) Experimental data and theoretical values for the isotope shifts $\Delta {\langle
      r^2 \rangle}_A= {\langle r^2 \rangle}_{A+2}-{\langle
      r^2 \rangle}_A $ for the even-even Pt isotopes (from
    \cite{hilberath92} and \cite{leblanc99}).}
  \label{fig-iso-shift}
\end{figure}

\begin{figure}[hbt]
  \centering
  \epsfig{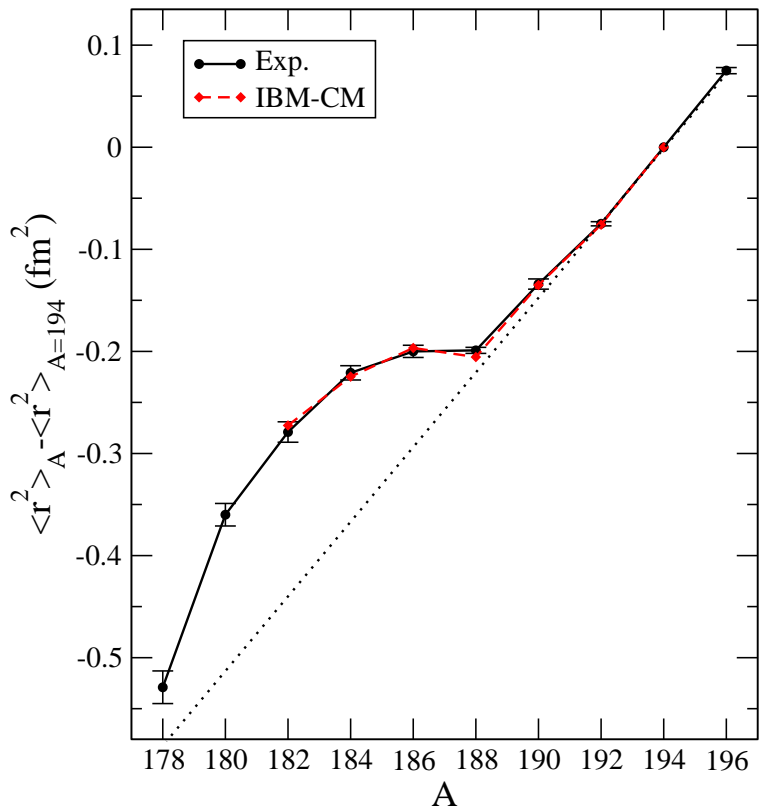} 
  \caption{(Color online) Experimental and theoretical variation of the mean-square charge radii 
    (relative to mass $A=194$). The dotted line
    is an extrapolation of the linear decrease for masses above
    $A=192$ (from \cite{hilberath92,leblanc99}).}
  \label{fig-iso-shift2}
\end{figure}

Experimental information about ground-state charge radii is also available
for both the even-even and odd-mass Pt nuclei.
Combined with similar data for the adjacent
Pb and Hg as well as the odd-mass Bi, Tl and Au nuclei the systematic
variation of the charge radii supplies invaluable
information on the ground-state wave function \cite{otten89}. In
particular, detailed studies by Hilberath {\it et al.}~\cite{hilberath92}
for the $^{183-198}$Pt nuclei and by Le Blanc {\it et
  al.}~\cite{leblanc99} 
have extended the charge radii measurements down to
$^{178}$Pt. We illustrate the relative changes defined as
$\Delta {\langle r^2 \rangle}_A \equiv \langle r^2 \rangle_{A+2}$ -$\langle
r^2 \rangle_{A}$ in Fig.~\ref{fig-iso-shift} and
the overall behaviour of ${\langle r^2 \rangle}_A$ relative
to the radius at mass $A=194$ in Fig.~\ref{fig-iso-shift2}. 
The mean-square charge radius exhibits a clear-cut change at and below mass $A=188$
with respect to the almost linear decrease for the heavier mass Pt nuclei, as can be seen in
Fig. \ref{fig-iso-shift2} .
This kink gives rise to a pronounced 
dip in the relative difference of charge radii for mass $A=186$ and $A=184$, as illustrated in 
Fig. \ref{fig-iso-shift}. 
An extrapolation of the linear trend downwards from mass $A=198$ (see 
dotted line in Fig.~\ref{fig-iso-shift2}),
hints towards an increased deformation of the $0^+$ ground state. 
This experimental mass dependence is rather well reproduced in the
Hartree-Fock-Bogliubov 
calculations (HFB) using the
Gogny force \cite{gogny80,gogny88}, as illustrated in Fig. 5 and
Table V in the study by Le Blanc {\it et al.}~\cite{leblanc99}.  

The IBM-CM calculations that were carried out by Harder {\it et al.}~\cite{harder97} as
well as the present, more detailed, IBM-CM study of the even-even Pt nuclei yield the same qualitative results 
for the decomposition of the ground-state $0^+$ wave function presented in Fig.~3 of \cite{Garc09}
and in the present Fig.~\ref{fig-composition}. 
The regular component with N bosons becomes minimal at $A\approx 182$ (about $10\%$
and $30\%$ in the more schematic and more extensive IBM-CM calculations, respectively) and reaches
a value of $80\%$ in both calculations for masses heavier than $A=188$ and lighter than $A=176$. This 
latter mass interval corresponds in a qualitative way to the bump in the evolution of the mean-square charge
radii relative to the dotted-line background. The mixing calculations carried out by Harder
{\it et al.} \cite{harder97} (see their Fig.~3) 
are consistent with a dip in the relative variation ${\Delta \langle r^2 \rangle}_A$
at the mass numbers $A=184$ and $186$. 
To calculate the isotope shifts, we have used the standard IBM-CM expression for the nuclear radius
\begin{equation}
r^2=r_c^2+ \hat{P}^{\dag}_{N}(\gamma_N \hat{N}+ \beta_N
\hat{n}_d)\hat{P}_{N} + 
\hat{P}^{\dag}_{N+2}(\gamma_{N+2} \hat{N}+ \beta_{N+2} \hat{n}_d) \hat{P}_{N+2}.
\label{ibm-r2}
\end{equation} 
The four parameters appearing in this expression are adjusted
to the experimental data. Note that only the
experimental values past mid shell ($A=182$) are used. The resulting values
are $\gamma_N=-0.099$ fm$^2$, $\beta_N=0.004$ fm$^2$,
$\gamma_{N+2}=-0.059$ fm$^2$, and $\beta_{N+2}=0.013$ fm$^2$ and are
only valid for the second half of the shell.  
The comparison with the experimental data show a very good
quantitative agreement, which confirms the assumption that
the balance between $[N]$ and $[N+2]$ contributions to
the wave function along the whole chain of Pt isotopes is very well described.


\section{Conclusions} 
\label{sec-conclu}
Upon comparison of the level systematics of the
Pb, Hg and Pt nuclei, from the neutron closed shell at
$N=126$ towards very neutron-deficient nuclei and even beyond the
neutron mid-shell at $N=104$, some conspicuous differences show
up. For the Pb and Hg nuclei, intruding bands are
observed in a compelling way and have been explained as the
occurrence of prolate and oblate bands (coexisting with the spherical states
at the $Z=82$ proton closed shell Pb nuclei) within the context of
mean-field theory or as many-particle many-hole proton excitations
across the $Z=82$ closed shell within a highly-truncated shell-model
approach that approximates the nucleon pairs as $s$- and $d$-boson pairs
(IBM).  For the Pt nuclei, however, the energy systematics does not obviously
point towards the presence of two different structures, as
was the case in the Hg nuclei.

In a former paper \cite{Garc09}, we have extensively compared
configuration mixing IBM calculations incorporating both 2p-2h
excitations $[N+2]$ and the regular configuration $[N]$ with IBM
calculations that restrict the model space to just the regular
configurations $[N]$. At first sight, one would expect to observe
strong differences. However, the results showed that, up to an
excitation energy of $\sim$ 1.5 MeV, the energy spectra, absolute
$B(E2)$ values, $B(E2)$ branching ratios, and quadrupole moments turned
out largely similar. The point was raised that, somehow,
configuration mixing did not show up explicitly when only considering a restricted set of data.
Therefore the name ``concealed'' is in order.

In the present paper, we have extensively studied how
configuration mixing between two distinct model spaces, i.e. the $[N]$
and $[N+2]$ configurations, may give rise to results that resemble those
obtained when only using a subset of the full model space. We have noticed
that it is important to have the two families of
energy bands ($i.e.$ the regular, $[N]$, and the intruder 2p-2h,
$[N+2]$, bands) of which the lowest cross at $A=176-178$ and
$A=186-188$. This particular crossing, reminiscent of similar
situations of inversion of regular and intruder configurations as
observed e.g. in the $N=20$ and $N=28$ neutron rich nuclei, and the
mixing between the regular and intruder 2p-2h bands gives rise to a
specific structure of the wave functions along the yrast bands. Near
mid-shell ($N=104$), we observe a progressive change of character from the
higher-spin members (at J$^{\pi}$=8$^+$,6$^+$) that are almost of pure
intruder character towards more mixed configurations, though still mainly of
intruder character, at the lower spin values. With the higher-spin
members retaining a rather pure intruder character for most of the mass region
studied here (with 172 $\leq$ A $\leq$ 192), it is natural to redraw
energy spectra relative to the higher-spin member at 8$^+$.  The
changing character in the wave function is evident from the energy
spectra, which result from the mixing of the regular configuration,
with a more spherical character and typical energy-spacing of $300-400$
keV, and the intruder 2p-2h configuration, with a more deformed character and
a typical energy spacing of $100-150$ keV. We have
illustrated this for the nuclei at $A=174, 180, 186$, and $A=192$, hence passing
through the mid-shell region. In addition to the study of energy spectra, we
have also carefully studied the decomposition of the most important E2
reduced matrix elements $\langle (f,J_f) \mid\mid \hat{T}(E2) \mid\mid
(i,J_i) \rangle$ into its components originating from the regular and
intruder bands. In this way, the specific effect of the mixing is
highlighted in both, the appearance of the correct energy spectra and
$B(E2)$ values, when comparing with the experimental data.

We stress in particular the importance of the crossing of these
unperturbed regular and intruder bands for the description of the specific
systematics of the energy spectra of the Pt nuclei. They are
characterised by a rather sudden drop in the excitation energy of the
0$^+_2$, 4$^+_1$, 2$^+_3$ and 6$^+_1$ levels between neutron number
$N=110$ and $N=108$, with energies starting to move up again between neutron number
$N=100$ and $N=98$. In the intermediate region, the energy spectra exhibit a
particularly flat behaviour with changing neutron number and even a
slightly 'upward' bump for the 0$^+_2$, 2$^+_2$, 4$^+_2$
levels. Within a schematic 2-level model, such an effect is caused 
by the mixing of a single regular band and an intruder band that has
parabolic-like evolution of the absolute energy. When one plots the
energy spectrum relative to the lowest 0$^+$ state, a slight bump results.

At this point, the remaining question is whether the configuration
mixing can be 'unveiled', in particular for the lowest-lying levels
such as the ground-state (through study of isotopic shifts, transfer
reaction intensities, etc) and lowest 2$^+_1$ state (g-factor for instance)
Therefore, we have calculated those
observables. The g-factor very clearly indicates the need for a rather
strong mixing, becoming more pure in regular $[N]$ character for the
lightest and heaviest mass numbers. Until present, transfer
reactions are not possible yet, but $\alpha$-decay can provide
such overlap factors through the hindrance factors. Even though not
quantitatively verified, the changing structure 
in the $[N]$ versus $[N+2]$ content of the wave functions for the 0$^+_{1,2}$
states is consistent with the change in hindrance factor, becoming increasingly
large for the excited 0$^+$ state compared to the ground state with decreasing mass
number. The isotopic shifts are also a very direct measure of the
ground-state wave function and as such is a number sensitive to its
precise decomposition. The dip in the isotopic shift curve at
$A=186$ is well reproduced by the present wave functions, derived from
the mixing calculations and containing two different structures, a more
spherical one and a more deformed component. The variation of the
mean-square radii clearly shows a bump structure very much centred
around the mid-shell $N=104$ neutron number.

The study of the Pt nuclei is interesting because it demonstrates
that calculations of a very different nature can give rise to a good description of a
number of properties. However, different models working in
different model spaces and with different effective interactions
should stand the test to as many observables as possible. In this
respect, the study of the configuration mixing is quite
illuminating as it consistently describes an as large set of observables as possible.
Ideally, one would like to see transfer data, populating
the Pt nuclei through single and double-nucleon transfer. Moreover, we suspect that
the Pt nuclei are not just an isolated case of concealed configuration
mixing. When carefully inspecting the
changing structure and systematics in the Po nuclei (which have two
protons outside the Pb core), the observed spectra do not display an obvious presence
of extra intruder bands. However, recent
studies point strongly towards the presence of intruding 2p-2h excitations (or
the presence of an oblate and a spherical band in mean-field terminology) near
$A=192$ \cite{hela99,coster99,oros99,grahn06,grahn08,grahn09,cocio11}.

\section{Acknowledgements}

We thank M.~Huyse, P.~Van Duppen and P.~Van Isacker for continuous interest in this
research topic and J.~L.~Wood for stimulating discussions on the
study of shape coexistence in the Pb region.
This work has been supported by Junta de Andaluc\'{\i}a under projects
FQM318 and P07-FQM-02962, by Spanish Consolider-Ingenio 2010 (CPAN
CSD2007-00042) and by the Belgian Interuniversity Attraction Pool
(IUAP) under project number P6/23. One of the authors (JEGR)
acknowledges the hospitality of the Department of Physics and
Astronomy of the University of Ghent. VH gratefully acknowledges a
postdoctoral fellowship from the F.R.S.-FNRS (Belgium) and the partial
financial support by the US DOE under grant DE-FG02-95ER-40934.  KH
thanks the FWO-Vlaanderen for financial support during this project
under grant G.0407.07N as well as the University of Ghent for
continuing support.

\end{document}